\documentclass[twocolumn,twoside,slac_two]{revtex4}
\usepackage{graphicx}
\usepackage{fancyhdr}
\usepackage[colorlinks,urlcolor=black,citecolor=black,linkcolor=blue]{hyperref}	
\usepackage{color, soul}	

\def\aap{{\it A\&A}~}

\def\apj{{\it ApJ}~}
\def\apjl{{\it ApJL}~}
\def\apjs{{\it ApJS}~}
\def\araa{{\it ARA\&A}~}
\def\mnras{{\it MNRAS}~}
\def\nat{{\it Nature}~}

\begin{document}

\title{Black Hole Jet Unification in the \emph{Fermi} Era}

%

\author{Rodrigo Nemmen}
\affiliation{NASA Goddard Space Flight Center, Greenbelt, MD 20771, USA}

\begin{abstract}
AGNs and GRBs produce powerful relativistic jets and their central engines share the same basic astrophysical ingredients, despite the vastly different mass scales. 
Using \emph{Fermi} and \emph{Swift} observations, we find evidence that the jets produced by blazars and GRBs follow the same correlation between the gamma-ray luminosity and kinetic power. This result suggests that jet production and energy dissipation mechanisms are remarkably similar over 10 orders of magnitude in jet power, establishing a physical analogy between AGN and GRBs. We discuss the implications of these results and the road ahead.
\end{abstract}

\maketitle

\thispagestyle{fancy}


\section{INTRODUCTION}

\begin{figure}
\centering
\includegraphics[width=\linewidth]{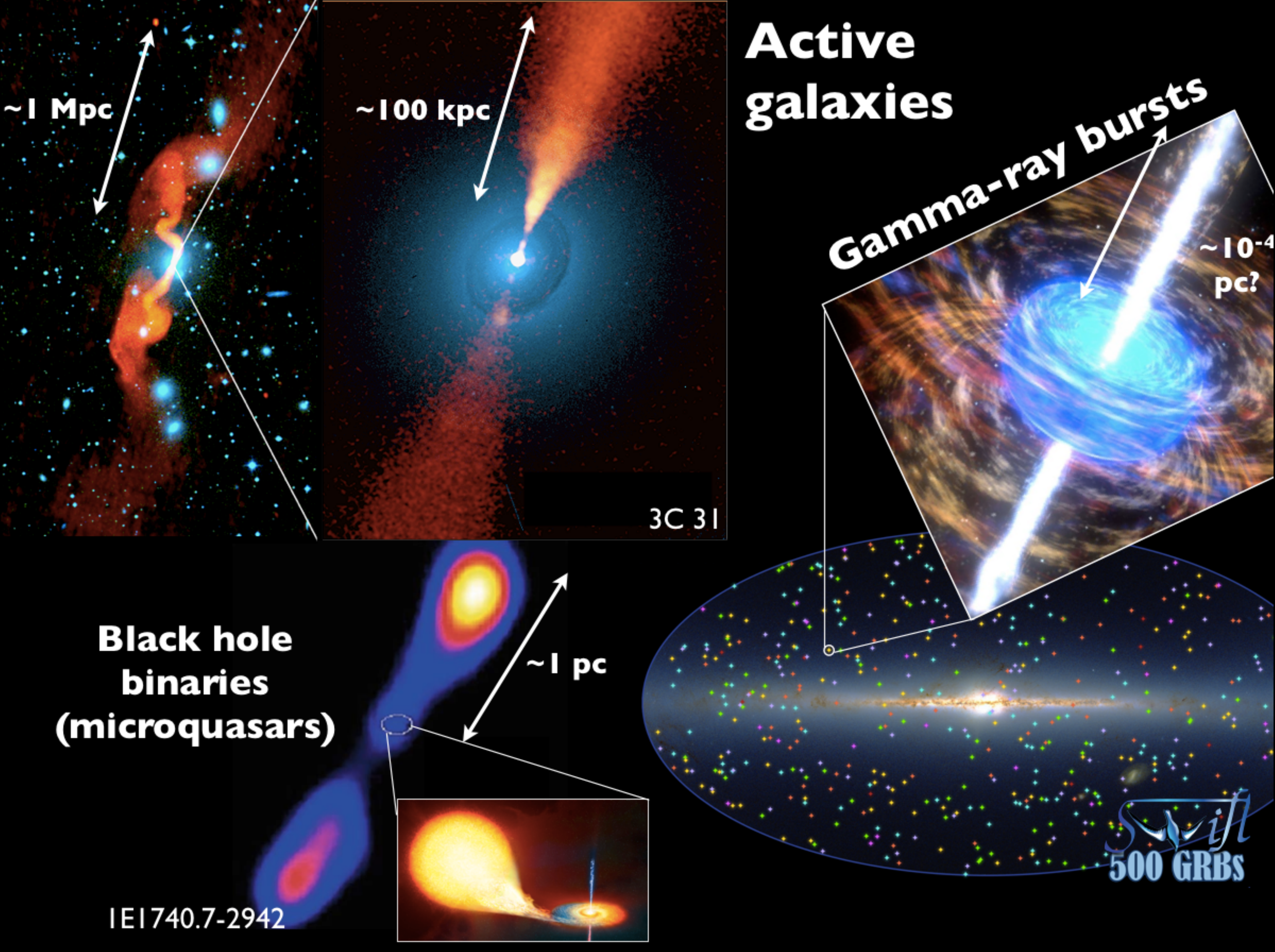}
\caption{Black holes caught in action producing relativistic jets. Mosaic based on: VLA and HST images \citep{Mirabel92,Martel99,Laing08} (\href{http://images.nrao.edu/257}{copyright NRAO}), GRB illustration (\href{http://www.nasa.gov/centers/goddard/news/topstory/2003/0618rosettaburst.html}{credit NASA/SkyWorks Digital}), \href{http://www.nasa.gov/mission_pages/swift/bursts/500th.html}{map of 500 GRB observed with \emph{Swift} until 2010 (credit: NASA/Swift/Francis Reddy)}, microquasar illustration (credit: ESA).}
\label{fig:mosaic}
\end{figure}

Relativistic jets are produced in different types of black hole systems in the Universe: active galactic nuclei (AGN), gamma-ray bursts (GRB) and black hole binaries/microquasars. Some beautiful examples of these systems are displayed in Figure \ref{fig:mosaic}. Despite decades of intense study and observations across all accessible wavelengths, many pieces of our understanding of jets are missing. For instance, we do not fully understand: (i) the mechanism responsible for their triggering: what is the interplay between black hole spin, accretion flow and magnetic fields? (ii) the nature of their energetics and high-energy emission; (iii) how does jet physics scale with mass from stellar to supermassive black holes?
Fortunately, a lot of progress has been done in these fronts \citep{Meier01rev,Narayan05,Ghisellini11rev,Sasha12rev,Abramowicz13}. Here, we will focus mainly on point iii listed above.

There is accumulating evidence that the black hole activity is similar in black hole binaries and AGNs (e.g.,  \citealt{Mirabel99,Marscher02,Merloni03,Falcke04, Mirabel04,Uttley05, Nipoti05,McHardy06,Kording06,Chatt11}). However, progress needs to be been done in assessing how the GRB phenomenon is related to radio-loud AGNs and microquasars \citep{Dermer99,Dermer07}. 

The same basic physical ingredients are involved in the process of jet production in radio-loud AGNs and GRBs (e.g., \citealt{Narayan01,McKinney06,Sasha08}). Therefore, we should expect some kind of scaling of the observed properties of the jet in both classes. However, such a connection remains elusive despite recent tantalizing results \citep{Wang11,Wu11}. 

Here, we present the results from an exploratory investigation of how jet physics scales with black hole mass \citep{Nemmen12}. We approach this issue from the observational side, making use of the multitude of high-energy observations of AGNs and GRBs made with \emph{Fermi}, \emph{Swift} and many other telescopes. 

\section{JET ENERGETICS IN BLAZARS AND GRBS}

As a first step in exploring connections between AGNs and GRBs, we focus on the energetics of the jets. For a proper comparison with GRBs, we restricted our AGN sample to include only blazars, such that all of our jetted sources are aligned with our line of sight. Our sample was selected due to the availability of:
\begin{itemize}
\item $\gamma$-ray luminosity $L^{\rm iso}$, which is a proxy of the jet bolometric luminosity
\item jet kinetic power $P_{\rm jet}$, estimated from the extended radio luminosities for the blazars whereas for the GRBs we relied on radio or X-ray afterglow measurements
\end{itemize}
The availability of these observables restricted our sample to 234 blazars [106 BL Lacs and 128 flat-spectrum radio quasars (FSRQs)] and 54 GRBs (49 long and 5 short GRBs, all with known redshifts $z$). 

\begin{figure}
\centering
\includegraphics[width=\linewidth,trim=20 0 50 40,clip=true]{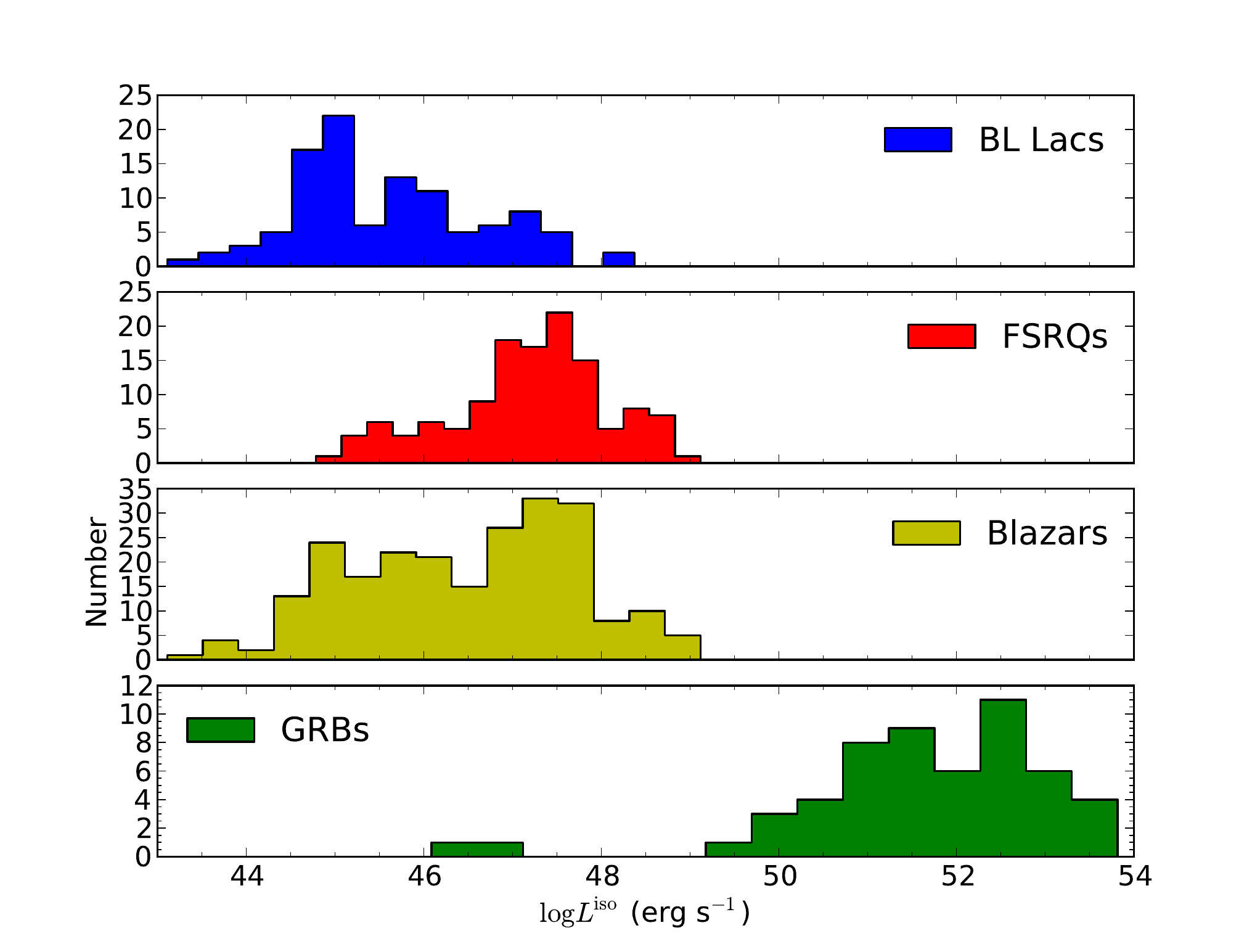}
\caption{Distribution of isotropic $\gamma$-ray luminosities for blazars and GRBs. Note the two low-luminosity GRBs -- 031203 and GRB 980425 -- which overlap with the blazar distribution.}
\label{fig:hist}
\end{figure}

For blazars, $L^{\rm iso}$ was estimated from the $\gamma$-ray energy flux and the spectral index measured with Fermi Large Area Telescope (LAT) \citep{2lac}. For GRBs, we define $L^{\rm iso} = E^{\rm iso}(1+z)/t_{90}$ where $t_{90}$ is the burst duration and $E^{\rm iso}$ is the isotropically equivalent energy radiated during the prompt emission phase and measured with different telescopes including Swift BAT, Fermi GBM/LAT, BeppoSAX, BATSE, HETE, HETE-2 and Integral. 
Figure \ref{fig:hist} shows the resulting distribution of these isotropic $\gamma$-ray luminosities for blazars and GRBs, illustrating the ten orders of magnitude range in luminosity of the sample.

\begin{figure}
\centering
\includegraphics[width=\linewidth,trim=20 0 50 40,clip=true]{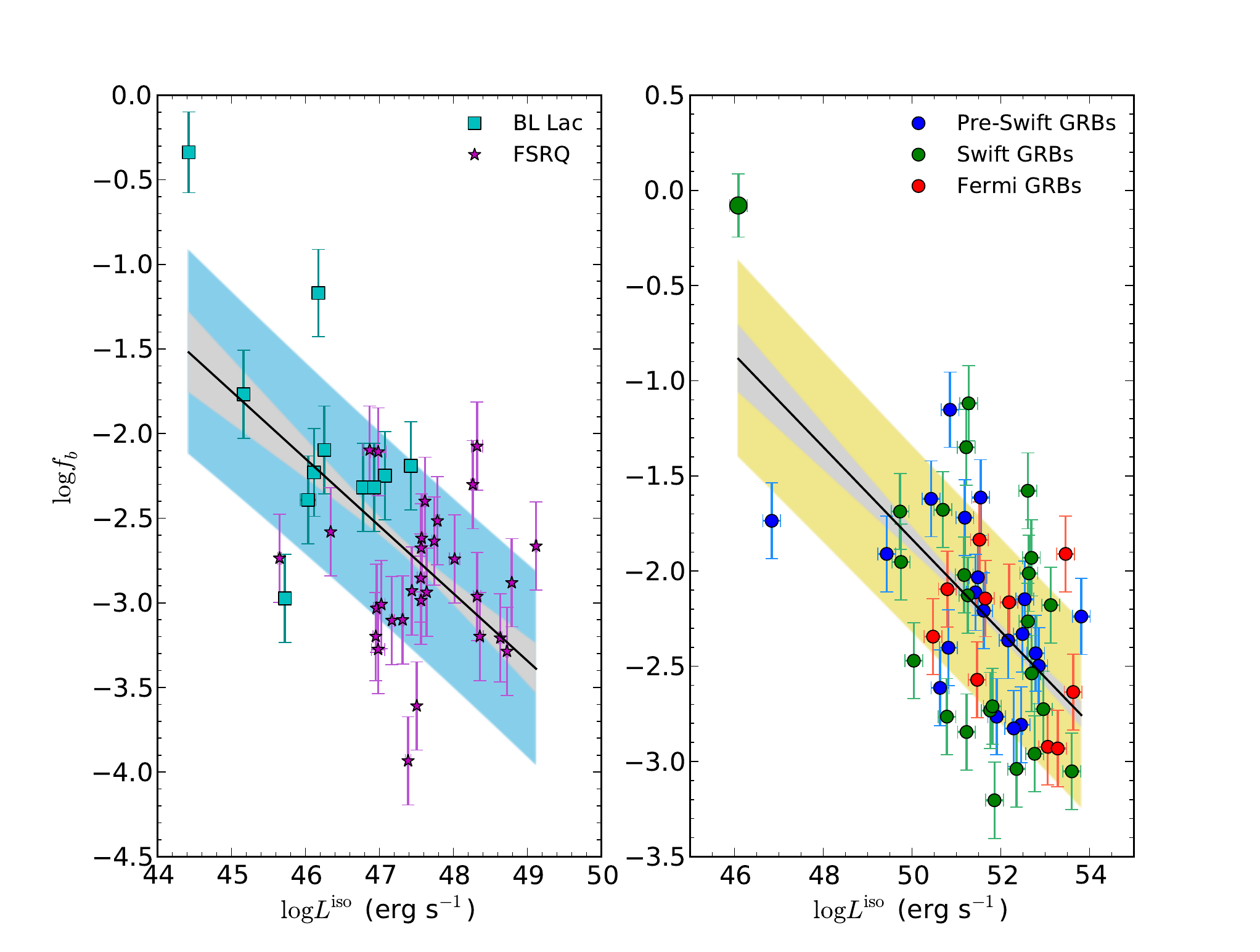}	
\caption{The relation between the apparent $\gamma$-ray luminosity and the beaming factor for blazars (left panel) and GRBs (right panel), significant at the $3.6\sigma$ and $4.4\sigma$ levels respectively. The solid lines correspond to the best-fit linear models obtained with a symmetric least-squares fit. The gray shaded region corresponds to the $1\sigma$ confidence band and the blue and yellow regions are the $1\sigma$ prediction bands. From \cite{Nemmen12}.}
\label{fig:beam}
\end{figure}

For an appropriate comparison of $\gamma$-ray luminosities and jet powers, what we really need to compare are the beaming-corrected luminosities $L=f_b L^{\rm iso}$ where $f_b$ is the beaming factor given by $f_b \equiv 1-\cos \theta$ where $\theta$ is the radiation cone half-opening angle. For GRBs, the beaming factor is computed from the jet opening angle $\theta_j$ as $1-\cos \theta_j$ estimated from the jet break in the GRB afterglow lightcurves \citep{frail01}; for blazars, $f_b$ is estimated as $1-\cos 1/\Gamma$ where $\Gamma$ is the bulk Lorentz factor of the flow. While an estimate of $\theta_j$ is available for each GRB in the sample, $\Gamma$ is only available for a subset of 41 blazars \citep{Pushkarev09}. Figure \ref{fig:beam} shows the anti correlation between $L^{\rm iso}$ and $f_b$ for both GRBs and blazars: \emph{brighter jets are more collimated and are affected by stronger beaming}.
Because $\theta$ is not available for the whole blazar sample, we used the power-law fit of $L^{\rm iso}$ vs $f_b$ displayed in Fig. \ref{fig:beam} as an estimator for $f_b$, taking into account the associated uncertainties.

The blazar kinetic power was estimated using the relation between the extended isotropic radio luminosity observed with the Very Large Array (VLA) and the jet kinetic power presented by \cite{Cavagnolo10}. For GRBs, we used $P_{\rm jet} = f_b E_k^{\rm iso}(1+z)/t_{90}$ where $E_k^{\rm iso}$ is the kinetic energy estimated from the radio (VLA) or X-ray luminosity during the afterglow phase using the standard afterglow model (e.g., \citealt{Racusin11}). These powers, as well the $\gamma$-ray luminosities, should be interpreted as values  averaged over the active state of jet production in AGNs and GRBs.

\section{COLLIMATION-CORRECTED ENERGETICS}

\begin{figure}
\centering
\includegraphics[width=\linewidth,trim=100 0 100 40,clip=true]{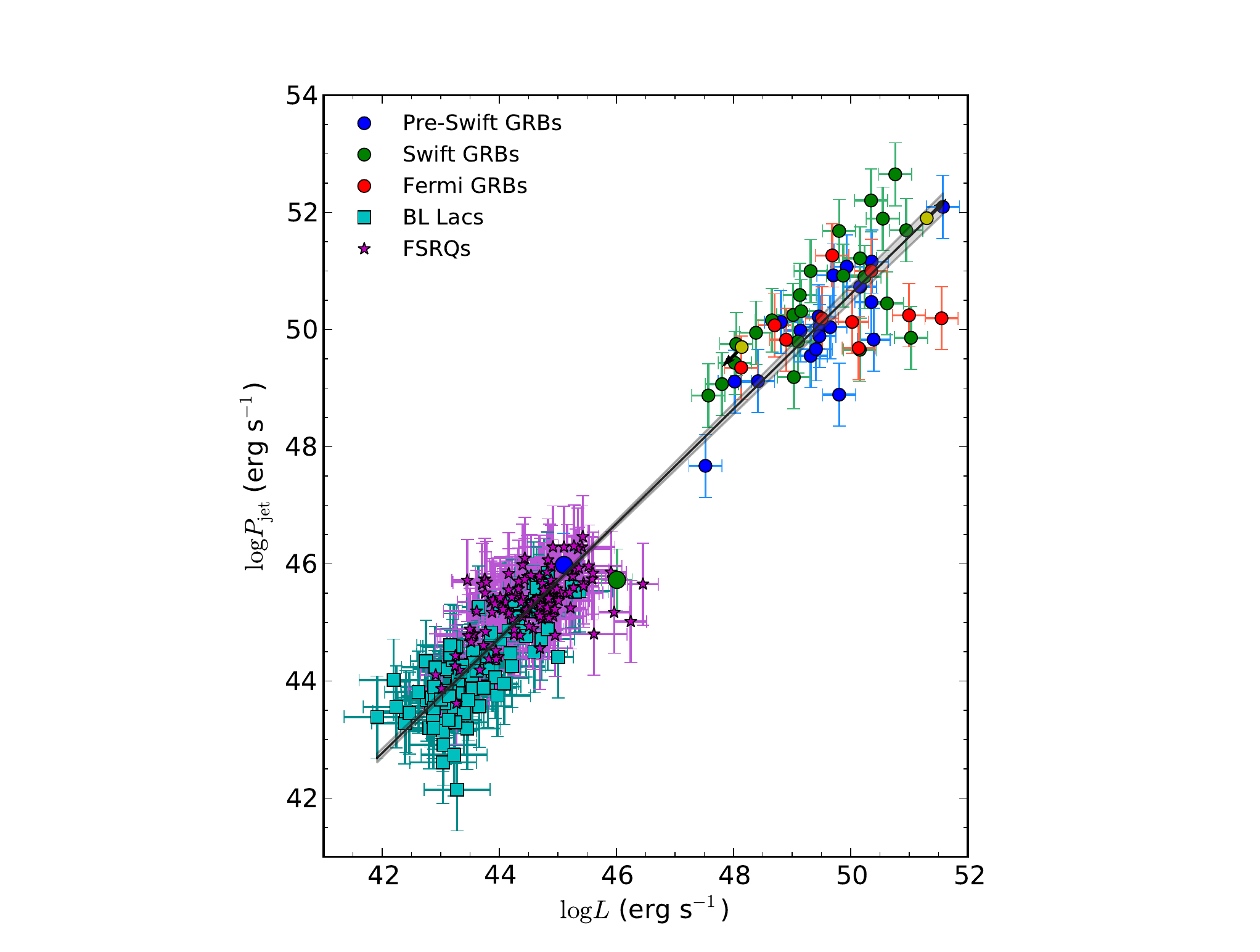}
\caption{The relation between the collimation-corrected $\gamma$-ray luminosity and the kinetic power for AGNs and GRBs. The shaded region displays the $2\sigma$ confidence band of the fit. The yellow data points correspond to XRF 020903 and GRB 090423 limits, which we do not take into account in the statistics. From \cite{Nemmen12}.}
\label{fig:main}
\end{figure}

Figure \ref{fig:main} displays the relation between the beaming-corrected $\gamma$-ray luminosity and jet power for blazars and GRBs. $L$ and $P_{\rm jet}$ are strongly correlated within the GRB and AGN samples. When studying luminosity-luminosity correlations, it is important to check possible spurious effects introduced by their dependence on the distance (e.g., \citealt{Merloni03}). We apply the partial Kendall $\tau$ correlation test \citep{Akritas96} and find that the probability for accepting the null hypothesis that there is no correlation between $L$ and $P_{\rm jet}$ is $\approx 10^{-12}$ and $10^{-9}$ for the blazar and GRB correlations, respectively. Therefore, the correlations are not a distance-driven artifact.

Blazars and GRBs follow very similar trends within the uncertainties. In fact, the whole blazar and GRB sample can be fit adequately with a power law over 10 orders of magnitude in luminosity \citep{Nemmen12}: 
\begin{equation}
P_{\rm jet} \approx 4.6 \times 10^{47} \left( \frac{L}{10^{47}} \right)^{0.98} \ {\rm erg \; s}^{-1}.
\end{equation}
In other words, ``Black hole engines'' produce relativistic jets that seem to maintain the same coupling between the total power carried by the jet and power radiated away: \emph{a universal scaling for the energetics of relativistic jets across the mass scale}. 

This scaling seems to be maintained regardless of the different environments and accretion flow conditions that lead to the production of powerful jets. For instance, radio-loud AGNs accrete at near-Eddington (FSRQs/FR II) or sub-Eddington (BL Lacs/FR I) rates (e.g., \citealt{Ghisellini09div}) whereas GRBs beat the Eddington limit by factors of $\gtrsim 10^{10}$. 

\section{RADIATIVE EFFICIENCY OF POWERFUL JETS}

\begin{figure}
\centering
\includegraphics[width=\linewidth]{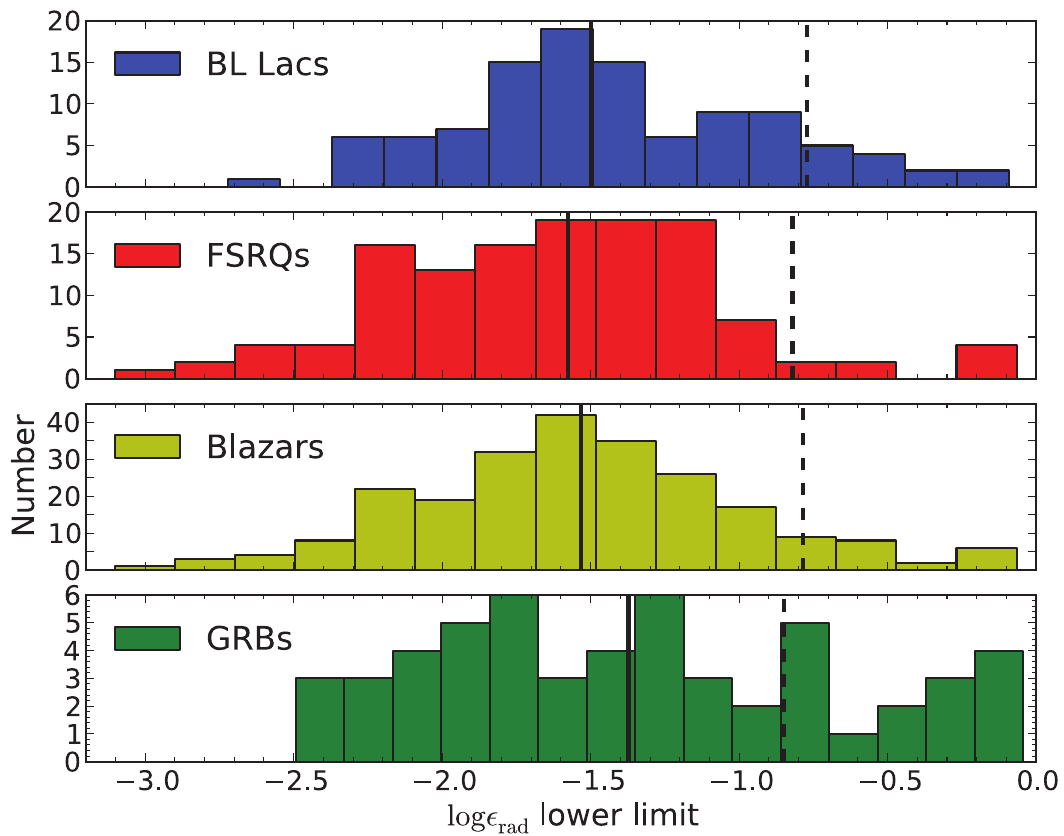}
\caption{Distribution of the $1\sigma$ lower limits on the radiative efficiencies for the jets in blazars and GRBs. The vertical solid lines indicate the median values of the lower limits and the dashed lines represent the median values of $\epsilon_{\rm rad}$ for each sample. From \cite{Nemmen12}.}
\label{fig:eff}
\end{figure}

We can also estimate the radiative efficiency of jets using the dataset we compiled. Defining the radiative efficiency in this context  as the fraction of the total jet power which is converted to $\gamma$-rays $\epsilon_{\rm rad} \equiv L/(L+P_{\rm jet})$, Figure \ref{fig:eff} shows the distribution of the lower limits on $\epsilon_{\rm rad}$ for AGNs and GRBs. 

Figure \ref{fig:eff} indicates that most of the jets in our sample dissipate at least $3\%$ of the power carried by the jet as radiation and overall they can radiate as much $15\%$. This range of efficiencies is considerably higher than previous estimates for AGNs based on radio to X-rays luminosities but they are in agreement with results obtained from blazar broadband spectral models as well as GRB afterglow studies (e.g., \citealt{Celotti08,Ghisellini10,Racusin11,Nemmen12}).

\section{CONCLUSIONS}

In summary, we found evidence that powerful relativistic jets in AGNs and GRBs follow a universal scaling for the energetics valid over ten orders of magnitude of luminosity, which is independent of the accretion physics and environment that lead to the production of these jets. Furthermore, we obtained a lower limit on the $\gamma$-ray radiative efficiency of these jets of $3\%$. The similarity in the energetics of blazars and GRBs suggests that there is a single fundamental mechanism responsible for the energy dissipation and production of relativistic jets in the Universe. 

A vast territory for the exploration of synergies between AGNs, GRBs and microquasars lies ahead, which will lead to further progress in understanding black hole (astro)physics. For instance, the energetics of jet production in microquasars and tidal disruption events (\citealt{Burrows11,Bloom11}; cf. Figure 1 in \citealt{Nemmen12}) certainly deserves investigation. Similarities in the spectral energy distributions of blazars and GRBs should  be systematically studied \citep{Wu11}. Studying the connection between the scaling in Figure \ref{fig:main} and the fundamental plane of black hole activity \citep{Merloni03,Falcke04} can lead to further insights on black hole jets.

Finally, we need to understand what constraints the energetics scaling sets on models for  shocks, particle acceleration and their dissipation efficiency in relativistic jets across the mass scale (e.g., \citealt{Nalewajko09,Sironi11,Nalewajko12,Yuan12grb}).

\bigskip 
\begin{acknowledgments}
The author wishes to thank: the organizers for such a productive symposium and for the opportunity of presenting his work, his collaborators as well as the \emph{Fermi} LAT Collaboration, and Terri Brandt for the great job in organizing the proceedings. The author was supported by an appointment to the NASA Postdoctoral Program at Goddard Space Flight Center, administered by Oak Ridge Associated Universities through a contract with NASA.
\end{acknowledgments}

\bigskip 

\end{document}